\documentclass[apj]{emulateapj}
\usepackage{natbib}
\usepackage{rotating}

\slugcomment{To appear in Astrophysical Journal}

\shorttitle{}
\shortauthors{Aravena et al.}

\begin{document}



\title{Unveiling the mask on the ULIRG-to-QSO transition object [H89]1821+643 at z=0.3: a gas-poor/gas-rich galaxy merger and the implications for CO-based dynamical mass estimates}


\author{M. Aravena\altaffilmark{1}, J. Wagg\altaffilmark{2}, P. P. Papadopoulos\altaffilmark{3}, I. J. Feain\altaffilmark{4}}




\altaffiltext{1}{National Radio Astronomy Observatory, 520 Edgemont Road, Charlottesville VA 22903\email{maravena@nrao.edu}}
\altaffiltext{2}{European Southern Observatory, Alonso de C\'ordoba 3107, Vitacura, Casilla 19001, Santiago 19, Chile}
\altaffiltext{3}{Argelander Institute for Astronomy, Auf dem H\"ugel 71, Bonn D-53121, Germany}
\altaffiltext{4}{Australia Telescope National Facility, CSIRO, P.O. Box 76, Epping, NSW 1710, Australia}


\begin{abstract}
We  report  the   detection  of  the  CO  $J=1-0$   emission  line  in [H89]1821+643, one of the most optically luminous QSOs in the local Universe and a  template ULIRG-to-QSO transition object, located in a rich cool-core  cluster at $z=0.297$.  The CO emission is likely to be extended, {\it highly asymmetric with respect to the center of the host elliptical where the  QSO resides}, and corresponds to a molecular gas  mass of  $\sim 8.0\times 10^{9}$\,M$_{\odot}$. The  dynamical  mass  enclosed  by  the CO  line  emission  region could amount up to $\sim1.7\times10^{12}$ M$_\odot$ (80\% of the total mass of  the elliptical host). The bulk of the CO emission is located at $\sim9$ kpc south-east from the nuclei position, and close to a faint optical structure, suggesting that the CO emission could either represent a gas-rich companion galaxy merging with the elliptical host or a tail-like structure reminiscent of a previous interaction. We argue that the first scenario is more likely given the large masses implied for the CO source, which would imply a highly asymmetric elliptical host. The close alignment between the CO emission major axis and the radio-plume suggests a possible role of the latter excitation of  the ambient  gas  reservoir. The stacking technique was used to  search for CO  emission and 3mm continuum  emission from galaxies in  the surrounding cluster, however no detection was found either toward  individual galaxies  or the stacked ensemble of galaxies, with  a  $3\sigma$ limit of $<1.1\times10^9$ M$_\odot$ for the molecular gas.
\end{abstract}


\keywords{galaxies clusters: general --- galaxies:formation --- galaxies: evolution --- galaxies individual (J1821+643) --- galaxies: ISM  --- galaxies:starburst --- galaxies: QSOs: general}



\section{Introduction}

In the general picture for QSO and stellar spheroid formation, two gas-rich disk galaxies merge and  the large amount of gas, compressed and  shocked by the interaction, trigger  intense starburst activity in the system, which is seen as a powerful IR luminous galaxy \citep[so  called Ultra  Luminous IR  Galaxies  -- ULIRGs;][]{Sanders1996}.   As the  system evolves, the gas flows into its central region, likely feeding an active galactic nuclei (AGN).  When the  black-hole has grown enough in size and luminosity, feedback and  gas   consumption  from   star-formation and AGN activity eventually  dissipates  the  dust  and an  optically  luminous  QSO emerges \citep{DiMatteo2005}. Once AGN activity subsides, the galaxy evolves into a  massive spheroid, hosting a quiescent  super-massive black-hole (SMBH) at its center. In  this scenario,  optically {\it and} IR luminous QSOs with warm IR colors ($S_{25\mu\mathrm{m}}/S_{60\mu\mathrm{m}}>0.2$) represent  the rare ULIRG-to-QSO transition objects  \citep{Sanders1988a,Sanders1988b}. 

Observations of CO line emission found large amounts of molecular gas  associated with these  ULIRG-to-QSO systems, in  support  of an evolutionary  scenario \citep[e.g.][]{Sanders1989, Young1991, Sanders1996}. However, due to the necessarily low resolution of single-dish telescopes, poor sensitivity of the few interferometric CO maps available and the lack of high-resolution optical images (e.g. with the HST) for most of these objects, few studies were able to determine the location of the molecular gas reservoir respect to the powerful QSO. \citet{Scoville2003}  imaged the CO line emission from a number of the most optically luminous PG QSOs, which are thought to be the end-product of the starburst-to-QSO transition. Interestingly, large amounts of gas, similar to that found in ULIRGs, where found in these objects. The conclusion drawn from these observations was that, assuming that the CO emission and starburst activity are well centered around the nuclei, the QSO host galaxies should be late-type spirals \citep{Scoville2003}. However, this contradicts the fact that most of these objects are typically found to have giant elliptical host galaxies in the optical. 

\begin{table*}[!ht]
\centering
\caption{Observed properties for the H1821+643 system}
\begin{tabular}{cccccccc}
\hline\hline
Source & $\alpha_\mathrm{CO}$ & $\delta_\mathrm{CO}$ & $z$ $^a$& $S_\mathrm{CO}$ $^b$& $S_\mathrm{CO}dv$ $^c$ & M$_\mathrm{gas}$ $^d$& $S_\mathrm{93 GHz}$ $^e$\\
              &           (J2000)  &   (J2000)      &  & (mJy beam$^{-1}$)   &    (Jy km s$^{-1}$)   &   ($\times10^{9}$ M$_\odot$)  & (mJy) \\
 \hline
H1821+643\ldots\ldots\ldots\ldots &  \ldots &       \ldots         &     0.2970     &     $<2.1$                     &    $<1.0$                  &   \dots     &   $10.3\pm0.5$   \\
Companion $^\dagger$\ldots\ldots\ldots\ldots  &  $18\ 21\ 57.42$ &    $+64\ 20\ 34.75$       &  0.2974          &         $2.1\pm0.5$                 &   $2.3\pm0.4$                  &  $(8.0\pm1.7)$  &  $<1.5$    \\
\hline
\end{tabular}
\begin{flushleft}
\indent $^\dagger$ CO emitting source.\\
\indent $^a$ redshift for H1821+643 nuclei taken from \citep{Schneider1992}\\
\indent $^b$ Peak emission measured in a CO image averaged over the line FWHM.\\
\indent $^c$ Velocity and area integrated CO emission.\\
\noindent $^d$ Gas mass computed from the measured CO integrated flux and using a X$_\mathrm{CO}$ prescription as explained in the text.\\
\noindent $^e$ Continuum flux density at 93 GHz.\\
\end{flushleft}
\end{table*}

Recently, sensitive high-resolution interferometric CO observations of a template ULIRG-to-QSO  transition system,  HE0450-2958  at $z\sim0.3$, along with Hubble Space Telescope (HST) imaging found a merger between a gas-rich spiral and a gas-poor elliptical, instead of   an optically luminous QSO hosted by a late-type gas-rich spiral in this system \citep{Papadopoulos2008}. Here, the optically luminous QSO lies {\it outside} the molecular gas reservoir, which is rather identified with a massive but obscured companion galaxy \citep{Papadopoulos2008, Elbaz2009}. High-resolution IR observations of this object support this assertion, and suggest also a more complex scenario where  the QSO radio jet  alters the state of  the ISM and possibly induces  a  starburst  in the  gas-rich  member of  the interaction   \citep{Klamer2004,Feain2007,Elbaz2009}.   Gas-poor/gas-rich  galaxy interactions ``igniting''  powerful  QSOs and starbursts are expected to be  common under the hierarchical framework  of structure formation \citep{Springel2005}.  Moreover,  an important consequence  of unrelaxed  dynamical merger configurations  between a  gas-rich system and a large QSO-hosting  elliptical would be the serious under-estimation of the dynamical mass of the  latter, an effect already noted for more orderly  (but  still unrelaxed)  gas-rich  spirals  at high  redshifts \citep{Daddi2010}. This  can have direct impact on CO-based estimates of QSO host galaxy masses at  high redshifts \citep{Wang2010} if they  involve such mergers.  Thus,  such reported ULIRG-to-QSO transition objects must be  revisited with high resolution CO and optical imaging,   in  order   to   reveal   the  detailed   dynamic configurations  of the molecular  gas reservoirs  with respect  to the AGN,  and investigate the possible  role of  the latter  in  the  formation  of  galaxies.  For  local  powerful  QSOs, typically  hosted by well-characterized  large ellipticals  with known masses \citep{Floyd2004}, high resolution CO observations can serve  as  excellent  test-beds  of  the accuracy  of  dynamical  mass estimates using CO lines.

One   of   the   most   prominent  local   templates   of   candidate ULIRG-to-QSO transition objects is IR/optically luminous QSO [HB89]1821+643  system  (hereafter  H1821+643)  at $z=0.297$. In  this paper, we report CO  emission line observations of this system using  the Combined  Array for  Research in  Millimeter-wave Astronomy (CARMA).  These observations directly  trace the bulk of molecular gas, which  is   the  fuel  for  star-formation.   We   use  a  concordance $\Lambda$CDM  cosmology   throughout,  with  $H_{0}=71$   km  s$^{-1}$ Mpc$^{-1}$,                $\Omega_\mathrm{M}=0.27$ and $\Omega_\mathrm{\Lambda}=0.73$ \citep{Spergel2007}.

\section{Observations}

\subsection{H1821+643}
With an optical absolute magnitude $M_\mathrm{V}=-27.1$  at a redshift of 0.297, H1821+643 is one of the most luminous QSOs in the local universe \citep{Hutchings1991, Schneider1992}. H1821+643 has been subject to a number of studies aimed at determining the morphology of its host galaxy. This bright QSO is hosted by a giant elliptical galaxy, with a half-light radius of $\sim14$ kpc in extent, as revealed by modeling of the point-spread function (PSF) subtracted HST images \citep{McLeod2001, Floyd2004}.  The luminous nuclear component outshines the underlying host with a luminosity ratio of $\sim11$ in the HST I-band \citep{Floyd2004}. H1821+643 lies in a massive galaxy cluster with optical richness class 2 \citep{Lacy1992}. X-ray observations show this corresponds to a strong cool-core cluster with a cooling time of 1 Gyr, a decrease of temperature in the cluster center from $9.0\pm1.5$ to $1.3\pm0.2$ keV, and a mass accretion rate from the central engine of 40 M$_\odot$ yr$^{-1}$ \citep{Russell2010}. 

Radio continuum imaging of this H1821+643 indicates this is a radio-quiet QSO, with a radio-lobe extending out to 250 kpc from the bright core in south-west direction \citep{Papadopoulos1995, Blundell1995, Blundell1996, Blundell2001}. The radio-lobe appears to bend by $\sim80^\circ$ in projection on sub-arcsec scales, suggesting that the radio axis is precessing, which implies a SMBH binary system at its nuclei \citep{Blundell2001} that has possibly already merged \citep{Robinson2010}. Interestingly, this object has radio luminosities at the transition to separate radio quiet and radio loud QSOs and also at the transition between FRI and FRII structures \citep{Blundell2001}.

\subsection{CO observations}

\begin{figure*}[!ht]
\centering
\includegraphics[scale=0.45]{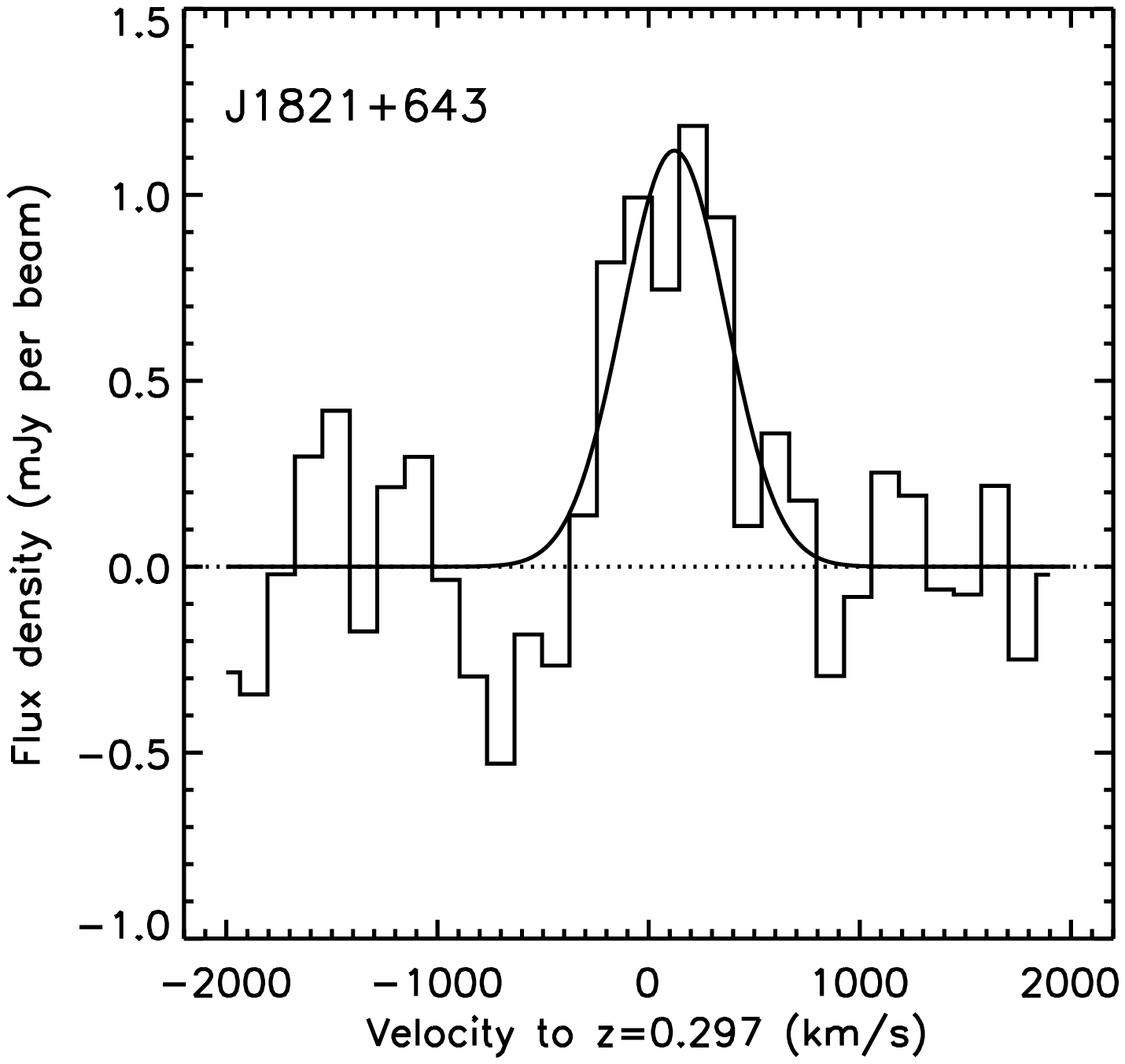}
\includegraphics[scale=0.45]{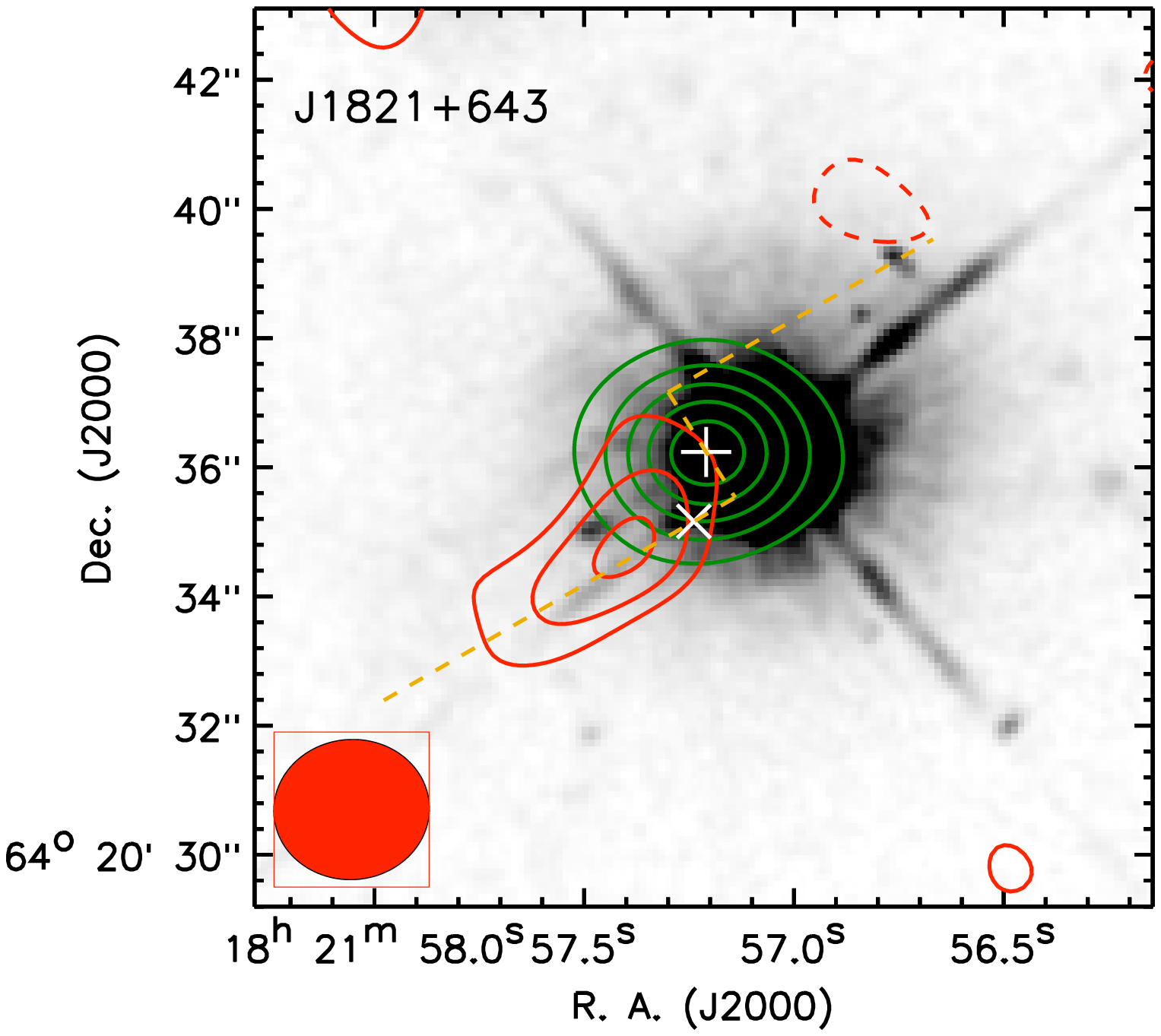}
\caption{Left: The HST I-band image is shown in the background with green and red contours overlaid showing the 3mm continuum and the CO 1-0 map integrated over 540 km s$^{-1}$, respectively, towards H1821+643.  Green contours are shown in steps of $8\sigma$, with $\sigma=0.2$ mJy beam$^{-1}$, starting at $\pm8\sigma$. Red contours levels are shown in steps of 1$\sigma=0.5$mJy beam$^{-1}$, starting at $\pm2\sigma$. The ``+'' and ``x'' white symbols represent the position of the bright radio nuclei and south-west radio component identified by \citet{Blundell1996}. The dashed orange line represents roughly the direction of the jet \citep[from][]{Blundell1996, Blundell2001}. The red ellipse at the bottom left shows the beam shape and size. Right: CO 1-0 spectrum of H1821+643 averaged over a region that encloses the CO emission shown in the left panel. The solid line represents a Gaussian fit to the data..\label{fig:1}}
\end{figure*}

Observations of the $^{12}$CO $J=1-0$ emission line (rest frequency: 115.271 GHz) were performed using CARMA in C-configuration with  two overlapping target fields. The first field was observed  during October to November 2008 centered at $\alpha_\mathrm{J2000}=18^h21^m57.21^s$, $\delta_\mathrm{J2000}=64^\circ20'36.23''$ (6 tracks; centered at the position of H1821+643). Observations of the second field were obtained during March 2010 in 2 tracks centered at $\alpha=18^h22^m02.8^s$, $\delta=64^\circ20'05.3''$, $\sim47.7''$ south-east of the first pointing. We offset the second pointing to allow us to simultaneously observe H1821+643 plus another source of interest in the field. This object will be studied in a separate publication.  The observations were taken under good millimeter weather conditions  with $\tau_\mathrm{230GHz}<0.5$. 

We used the 3 mm receivers tuned to 88.8751 GHz (lower sideband). Three bands (or spectral windows) of 15 channels each and 31.25 MHz per channel were used. These bands were put together to roughly cover $\sim1.3$ GHz in frequency. For the 2010 observing runs, the three bands were overlapped by 2 channels in order to improve the bandpass calibration, and the edge channels were flagged. In the 2008 runs, however, the bands were not overlapped and thus no flagging of the edge channels could be performed as it would translate in spectral gaps. This resulted in a loss of sensitivity in channels with velocities of $\sim\pm780$ km s$^{-1}$. After combining all datasets, the effective bandwidth was $\sim1.281$GHz. The large bandwidth allowed us to search for emission not only from H1821+643 but also from galaxies in its dense environment. The upper sideband was tuned to $\sim93.1$ GHz to obtain an independent measurement of the continuum emission. 

We observed the nearby source J1849$+$670 (distance to target $\sim$3.9$^\circ$) every 15 min for amplitude and phase calibration. The strong calibrators J1927+739, 3C273 and 3C454.3 were used for bandpass calibration. Neptune and MWC349 were used for absolute flux calibration.  Pointing was checked every hour using the optical and radio methods \citep{Corder2010}. The data were edited and calibrated using a combination of MIRIAD and CASA tasks. The first and last edge channels where the bandpass deteriorated were flagged accordingly. The large primary beam of CARMA at the observing frequency ($1'$) allowed us to partially overlap the two pointings resulting in a 20\% increase in sensitivity. We subtracted the continuum emission from our data in the $uv-$plane by fitting the continuum with a polynomial of order 1, using line-free channels to estimate the continuum level. This is particularly important for the H1821+643 line emission, for which strong continuum from the jet is expected (see below). 

The naturally weighted visibilities were imaged and deconvolved using CLEAN to a residual of approximately 1$\sigma$, where $\sigma$ is the noise level, in a box centered in H1821+643. The final images reach a sensitivity of $1\sigma\approx0.7$ mJy per 120 km s$^{-1}$ channel in each field, with a clean beam size of $2.5''\times2.1''$ and a position angle of $95^\circ$.

\section{Results}

\subsection{CO and continuum emission in H1821+643}

Figure \ref{fig:1} shows the CO velocity-integrated emission map and the corresponding spectrum averaged over the emitting region towards H1821+643. A Gaussian fit to the spectrum indicates an emission line with a full-width at half maximum (FWHM) of $580\pm140$ km s$^{-1}$ and centered at $120\pm60$ km s$^{-1}$ from the optical redshift $z=0.297$ \citep{Schneider1992}. By fitting an elliptical Gaussian profile to the integrated CO map, we find that the emission appears to be extended along the major axis with a  scale of $4.0''\pm1.2''$ or $18\pm5$ kpc, however it is unresolved along the minor axis (Position Angle $132^\circ\pm8^\circ$). By fixing the extent of the emission to that value, we estimate an integrated CO emission of 2.3$\pm$0.5 Jy km s$^{-1}$. The source is centered at $\alpha=18^h21^m57.43^s$ and $\delta=64^\circ20'34''.80$, with a positional uncertainty from the fit of $0''.8$, in agreement with the expected positional uncertainty given by $\Delta \theta=0.5\times\mathrm{SNR}\times\theta_\mathrm{beam}^{-1}=1.1''$, where SNR$\sim5$ corresponds to the signal to noise ratio and $\theta_\mathrm{beam}\sim2.2''$ is the half-power beam width (HPBW).

H1821+643 has an integrated continuum emission of $10.9\pm0.7$ mJy and $10.3\pm0.5$ mJy at 88.9 GHz and 93.1 GHz, respectively. The source is centered at  $\alpha=18^h21^m57.207^s$ and $\delta=64^\circ20'36.''22$, with an uncertainty in the fit of $0.06''$.  The emission is slightly resolved with a deconvolved size of$\sim0.4'' - 1.0''$ or $1.5 - 4.3$ kpc. As shown in Fig. \ref{fig:1}, the continuum emission can be easily identified with the location of the radio core of this QSO \citep{Blundell1996}.

\subsection{Previous reported CO observations of H1821+643}

Previous searches for CO in this object with the IRAM 30m telescope failed detect any emission \citep{Alloin1992, Combes2011}. \citet{Combes2011} reports a $3\sigma$ limit of 1.7 Jy km s$^{-1}$ in a 300 km s$^{-1}$ channel. In order to directly compare this upper limit with our observations, we deconvolved the visibilities to a coarser velocity resolution of 300 km s$^{-1}$, matching the one used by Combes et al., around the central velocity of the line profile. Integrating spatially over the emission area and over the velocity range covered, we find an integrated flux density of $\sim0.72$ Jy km s$^{-1}$ over a 300 km s$^{-1}$ channel, which is about 2.3 times deeper that Combes et al. observations. Hence, this explains why the CO emission line was undetected previously.

\subsection{Gas off the nuclei: a gas-rich/gas-poor galaxy merger or tail-like structure?}
\begin{figure}[!t]
\centering
\includegraphics[scale=0.6]{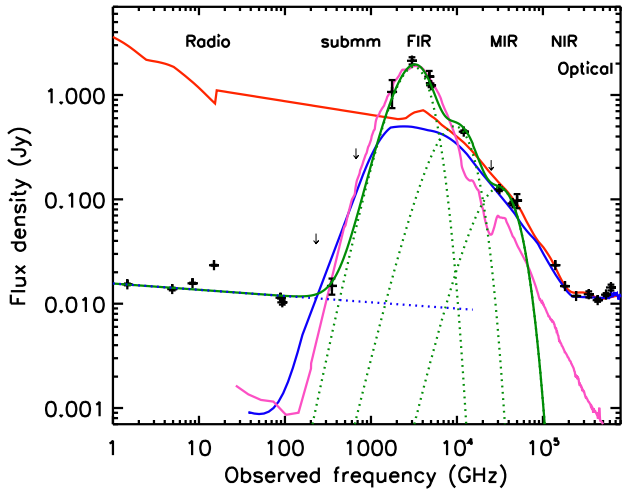}
\caption{Spectral energy distribution of H1821+643. The black crosses show the data points from radio to optical wavelengths. The green and blue dotted lines represent the dust and radio emission models fitted to the data, with the solid green line showing the combined model. The red and blue solid lines show the redshifted radio loud and radio quiet QSO template SEDs from \citet{Elvis1994} and \citet{Polletta2007}, respectively, scaled to match the optical emission. The magenta solid line shows the redshifted template SED of MrK231, a typical starburst to QSO transition object.\label{fig:sed}}
\end{figure}

The CO emission line from H1821+643 is located $\sim2.0''$ (about 1 synthesized beam; $\sim8.8$ kpc) south-east of the radio-core position, and $2.8''$ ($\sim12.3$ kpc) south-east of the optical center (host galaxy). Based on analysis of PSF subtracted HST optical images, \citet{Floyd2004} find a nebulous optical structure. Floyd et al., however, point out that such optical feature may correspond to an artifact of the PSF subtraction procedure.

The CO position coincides with the location of this optical structure, giving support to the reality of such feature. Indeed, the bulk of the molecular gas is located at only $0.52''$, well within the uncertainty in the position, from an optically faint peak to its north-east (Fig. \ref{fig:1}). Thus, this suggest that the CO emission is possibly associated with either an optically faint but gas-rich companion galaxy that is merging with the giant elliptical host, or with an asymmetric tail-like structure within the host galaxy reminiscent of a previous merger event. Note, however, that the powerful AGN is hosted by a giant elliptical galaxy, and large amounts of molecular gas (M(H$_2$)$>5\times10^9$ M$_\odot$) such as that observed here (see below) are rare in early-type galaxies \citep{Crocker2011}. Only 3 out of 56 early-type galaxies in the Virgo cluster have molecular gas masses $>10^9$ M$_\odot$ \citep{Young2011}, being statistically unlikely that the molecular gas observed in H1821+643 is related to the host galaxy. Although it is still plausible that  the gas could be related to a tail, relic of a past merger, the substantial amounts of molecular gas, its offset location with respect to the elliptical host and the existence of a starburst component in the optical-to-radio SED (see below) suggest that the CO emission actually comes from a different galaxy. If the molecular gas is indeed related to the host galaxy, it would imply a highly asymmetric gas distribution, possibly associated with a previous interaction or spiral arm, in an unusually gas-rich elliptical galaxy, likely related with starburst activity. Deep high-resolution IR observations, transparent to obscuration, are necessary to decide between both possibilities.

The CO emission is also well aligned with the radio-axis to the south-east suggesting a physical association between the radio-plume emission, the diffuse optical source and the large amounts of molecular gas. According to \citet{Blundell2001}, the radio jet appears to have been pointed within $7^\circ$ within the last 0.3 Gyr. Thus, the probability that the CO emission is by chance located within 7$^\circ$ is $\sim2\times7/360$ or 3.9\%, where the factor 2 accounts for both radio lobes (north and south). Such probability is similar to the one found based on the alignment between the radio-axis and the CO position angle. 

\subsection{ISM properties} 

The CO luminosity (in units of K km s$^{-1}$ pc$^2$) can be derived from the line intensity following \citet{Solomon1997} as,
\begin{equation}
L'_\mathrm{CO} = 3.25\times10^7 \ \nu_\mathrm{obs}^{-2}\ (1+z)^{-3}\ D_\mathrm{L}^2
\int_{\Delta V} S_\mathrm{CO} d v\
\end{equation}
where $\int_{\Delta V} S_\mathrm{CO}d v$ is the velocity integrated CO flux density in units of Jy km s$^{-1}$, $\nu_\mathrm{obs}$ is the observing frequency in GHz, and $D_\mathrm{L}$ is the luminosity distance in Mpc at the source redshift $z$. From our integrated CO flux, we derive $L'_\mathrm{CO}=(1.0\pm0.2)\times10^{10}$ (K km s$^{-1}$ pc$^2$) . The luminosity of the CO $1-0$ emission line is commonly used to estimate the molecular gas mass, M(H$_2$), by using the relation M(H$_2$)=$X_\mathrm{CO} L'_\mathrm{CO}$. $X_\mathrm{CO}$ is the CO luminosity to gas mass conversion factor which we assume to be 0.8 M$_{\odot}$  (K km s$^{-1}$ pc$^2$)$^{-1}$, as it was found for local IR luminous galaxies \citep{Downes1998}. We thus estimate a molecular gas mass of $(8.0\pm1.7)\times10^{9}$ M$_\odot$ for the CO emitting source. This molecular gas mass is at the high end of the mass range found for local ULIRGs, and is typical of luminous submillimeter galaxies and massive disk galaxies at high-redshift \citep{Greve2005,Solomon2005,Daddi2010}. 

\subsection{Dynamical mass of the CO-bright source}

Due to the relatively low significance of the CO detection, it is not possible to determine the actual geometry of the CO emitting source. We estimate the dynamical mass based on the CO profile using disk geometry, which is commonly assumed for CO sources. However, we remark that our adoption of a disk geometry is not well established for our CO source and we use it only as a guide and for the purpose of comparison with other CO studies. In this case, the dynamical mass is given approximately by M$_\mathrm{dyn}\approx (R/G)(v_\mathrm{rot}/sin(i))^{2}$. The source radius can be expressed in terms of its major axis, as  $R\sim a_\mathrm{major}/2=9$ kpc, and the rotational velocity is given by the CO line-width $\sim580$ km s$^{-1}$. The  CO  emission source has a major  axis  at  the  redshift   of  the  source of $\sim18\pm5$  kpc (at the $3\sigma$ level) and a minor axis limited  to the beam size ($\lesssim10$ kpc). This allows us to put a limit on the inclination angle given by  $i=\mathrm{cos}^{-1}(a_\mathrm{minor}/a_\mathrm{major})>40^\circ$. This takes into account the uncertainty in the major axis and the value for the inclination angle would only increase with smaller minor axis. With these values, we find a dynamical mass of $<1.7\times10^{12}$ M$_\odot$.

The  observed  CO  distribution is  highly  asymmetric  with respect to  the QSO  position itself and  actually barely  includes it (see  Figure 1).  Based on the accurate optical photometry of H1821+643 from HST images, \citep{Floyd2004} estimates a total mass for the elliptical host galaxy of $2.1\times10^{12}$ M$_\odot$. Given the asymmetric distribution of CO-emitting gas close to the QSO host galaxy, it is indeed surprising that  the dynamical mass range  estimated for the CO  source amounts up to $\lesssim80\%$ of the mass of the elliptical host ($\sim2.1\times10^{12}$ M$_\odot$). As explained above, such large amounts of molecular gas are rare (but possible) in old giant ellipticals \citep{Crocker2011, Young2011}, and thus the CO emission could potentially arise from an obscured, merging starburst galaxy, which in this case would have an important stellar component. On the other hand, it is interesting to note that if the CO profile would have been used to estimate the QSO host galaxy dynamical mass, we would have underestimated such value by at least 20\%. This is consistent with results from high-redshift disk galaxies, for which the CO-based dynamical mass was found to be underestimated by $\sim30\%$ \citep{Daddi2010}.


The potential discrepancy  between M$_{\rm dyn}$(CO) and $\rm M_{QSO-host}$  in massive mergers  offers  a  general cautionary tale  when it comes to CO-deduced  dynamical mass estimates of  QSO host  galaxies at  the higher redshifts \citep[e.g.][]{Wang2010}. 

\subsection{Spectral energy distribution}

\begin{figure}[!t]
\centering
\includegraphics[scale=0.45]{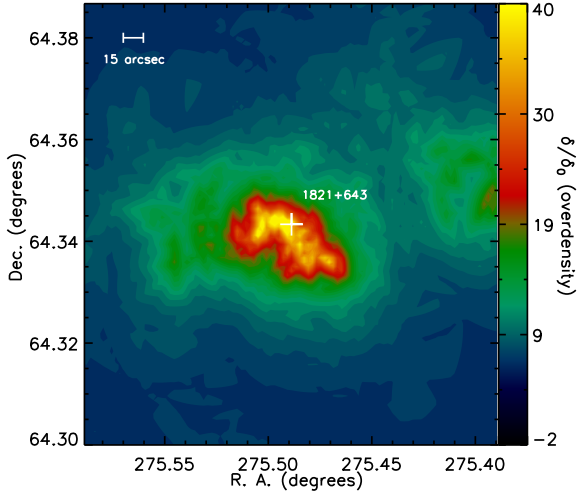}
\caption{Density of galaxies in the redshift slice $z\sim0.25-0.35$ in the H1821+643 field. The contour levels indicate the overdensity of galaxies with respect to the average number of galaxies found in a control field to the south.\label{fig:density}}
\end{figure}

Using radiative transfer models of starburst and AGN emission, \citet{Farrah2002} studied the spectral energy distribution (SED) of H1821+643. These models give a good fit to the far-IR data points although they fail to describe the emission from old stars at wavelengths shorter than 4 $\mu$m. Their best fit model to the IR emission of H1821+643 indicates a composite object, with a warm AGN component that provides about 60\% of the total luminosity, and a starburst component with a cold dust temperature of $\sim30$ K and a dust mass of $\sim3\times10^8$ M$_\odot$. Their model, however, does not make reference to the radio emission, which could contribute a significant amount to the emission at submillimeter wavelengths. 

The spatial coincidence between the radio core \citep{Blundell1996} and the peak emission in our 93 GHz continuum images of H1821+643 implies that most of this emission comes from the nuclei and that it is non-thermal. As shown in Figure \ref{fig:sed}, the flux density at $\sim93$ GHz is compatible with the trend followed by the radio emission at lower frequencies up to 350 GHz, in support for the non-thermal nature of such emission. 

In order to account for the radio continuum emission, a power-law parametrization was used, of the form $S_\nu \propto (\nu/\nu_0)^\alpha$, while simultaneously modeling the far-IR emission with a multi-component gray-body spectrum . Each dust component can be described by a model of the form $S_\nu \propto B_\nu(T_\mathrm{d})(1-e^{-\tau_\nu})$, where $B_\nu(T_\mathrm{d})$ is the Planck function and $T_\mathrm{d}$ is the dust temperature. The dust optical depth,  $\tau_\nu$, is proportional to the dust mass, $M_\mathrm{d}$, and to the dust absorption coefficient $\kappa(\nu)=(\nu/\nu_0)^\beta$, which we adopt to be equal to 0.04 cm$^2$ gr$^{-1}$ at 250 GHz \citep{Kruegel1994} with $\beta=2.0$ \citep{Priddey2001}.

The best fit model is shown in Figure \ref{fig:sed}. A cold dust component is found with $T_\mathrm{d}=50$ K and a dust mass of $1.4\times10^8$ M$_\odot$. A second, warmer dust component was obtained with a dust temperature of 130 K, and a dust mass of $\sim1.0\times10^6$ M$_\odot$. Both components alone do not allow us to fit the shorter wavelengths data points, $\lambda=6-10\ \mu$m , and thus we added a third, hot component with a dust temperature of $430$ K, which we expect to be biased by the unknown amount of emission from stars at these wavelengths. 

The radio emission is well described by a power-law with a slope of $\alpha\sim-0.05$ and a flux density of $16$ mJy at 1.0 GHz. The dust model obtained is consistent with that found by \citet{Farrah2002}, however the inclusion of the radio emission implied a change in the resulting cold gas temperature from 30 K to 50 K. 

The cold dust component emission is identified with vigorous starburst activity, possibly related to the CO emitting source, and the warm/hot dust component emission with AGN activity. We find a total IR luminosity $L_{\mathrm{8-1000}\mu\mathrm{m}}=1.0\times10^{13}$ L$_\odot$, with the AGN contributing with  $5.3\times10^{12}$ L$_\odot$ and the starburst with $4.7\times10^{12}$ L$_\odot$. Note that although the flat spectrum radio emission contributes with about 1\% of the total IR luminosity, it amounts to $\sim70\%$ of the emission at 350 GHz. On the other hand, the starburst emission is negligible at 90 GHz, $\sim25\ \mu$Jy, consistent with our observations. Finally, note that the dust mass for the cold dust component and our CO measurements imply a gas to dust mass ratio of $\sim60$, comparable to the values found for a sample of local far-IR selected galaxies \citep{Dunne2000, Seaquist2004} and distant submillimeter galaxies \citep{Greve2005, Kovacs2006, Michalowski2010}, but somewhat low compared to values found for ULIRGs \citep{Wilson2008}.

\section{Discussion} 

\subsection{Another possible example of jet-induced galaxy formation}

\begin{figure}[!t]
\centering
\includegraphics[scale=0.45]{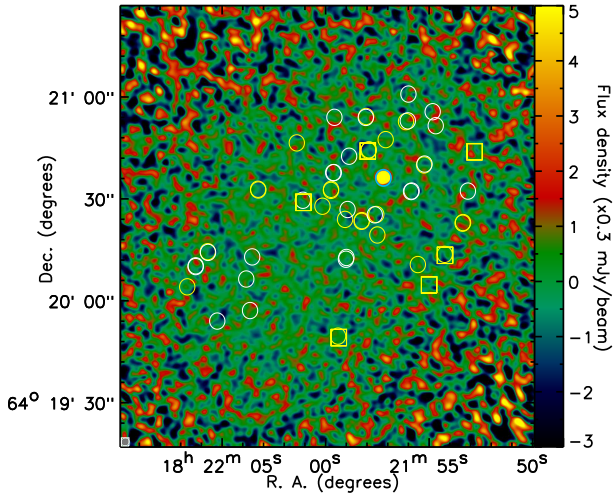}
\caption{Mosaic of the 93 GHz continuum emission obtained by combining the data obtained in the two overlapping target fields. The flux density scale in terms of the central noise level (0.3 mJy) is shown to the right. The cyan circle represents the position of H1821+643. Yellow squares show the position of the galaxies with redshifts matching the frequency coverage of the CO line in the cluster ($z=0.2925-0.3015$). Yellow and white circles represent galaxies with photometric redshifts in the range $z=0.25-0.35$ (24 galaxies) and outside that range (25 galaxies), respectively. The size of circles and boxes is $2\times\theta_\mathrm{beam}$.\label{fig:mosaic}}
\end{figure}

Simulations predict that galaxies receive a large fraction of their gas through cold flows of intergalactic gas. Such neutral gas is found in filaments and cools effectively in massive galaxy clusters onto giant elliptical galaxies at their centers \citep{Fabian1994}. Feedback from the central radio source may provide a mechanism for reheating the gas. It has been proposed that the cooled gas in the inner regions of the cluster may converge toward the central black-hole or collapse to form molecular gas clouds and stars \citep{Pizzolato2005, Russell2010}. CO observations of the central massive galaxies in clusters indicate that the deposition of gas toward the central regions is indeed produced in short timescales \citep{Edge2001, Salome2003, Edge2003}. 

It is well-known that H1821+643 lies at the center of a massive cluster of galaxies \citep{Lacy1992} and therefore the cold accretion scenario in this case is possible. The gas funneled to the center would feed the massive black-hole, and an accretion disk instability would ignite the jet emission \citep{Russell2010}. However, to start the outburst a galaxy merger scenario is more plausible since it would explain the precessing jet axis, which is likely produced by a binary black-hole system \citep{Russell2010}. Our observations suggest that the actual answer is more likely a combination of both hypotheses. The CO emission is offset from the QSO nuclei and coincident with the position of a faint optical feature to the south-east and well-aligned (position and position angle) with the southern jet axis. 

X-ray observations indicate that the cluster around H1821+643 has a cooling rate of $300\pm100$ M$_\sun$ yr$^{-1}$, whereas the QSO accretion rate, 40 M$_\odot$ yr$^{-1}$, is at half the Eddington limit \citep{Russell2010}. This implies that $\sim3\times10^{11}$ M$_\odot$ has cooled down in the typical timescale of 1 Gyr of this cluster and is potentially available for molecular gas formation and material accretion onto the central black-hole. At this accretion rate, the QSO would only keep growing for a fairly short period of time, $\lesssim0.1$ Gyr, since in this time it would grow larger than the upper limit that has been observed in the local Universe \citep{Russell2010}. This puts a limit on the period in which the QSO is active and might have formed jets, and is consistent with the timescale of $0.3$ Gyr in which the QSO jet appears to have remained within a 7$^\circ$ cone \citep{Blundell2001}. This implies that the jet axis has been pointed in fairly the same direction during most of its existence. 

The significant alignment  between the jet axis and the companion galaxy  as well as the jet lifetime (the time  in which the jet has  been impacting the  merging galaxy) suggest that  the jet could have a role in boosting star formation within the CO emitting source \citep{Klamer2004, Feain2007, Elbaz2009}.  Strong jet-ISM  interactions  inducing  highly-excited  CO lines  over  large molecular  gas  reservoirs  has  been  recently noted  in  3C\,293,  a powerful FR\,II  radio galaxy \citep{Papadopoulos2008b},  though its effect  on  the SFR  (very  low in  3C\,293)  remains  far from  clear \citep{Papadopoulos2010}. Examples of radio-loud QSOs where the jet aligns with the location of the bulk of molecular gas have been found at high-redshift \citep{Klamer2004, Elbaz2009}.


\subsection{Searching for cold flows of gas as the origin for CO abundance}

\begin{figure*}[!t]
\centering
\includegraphics[scale=0.45]{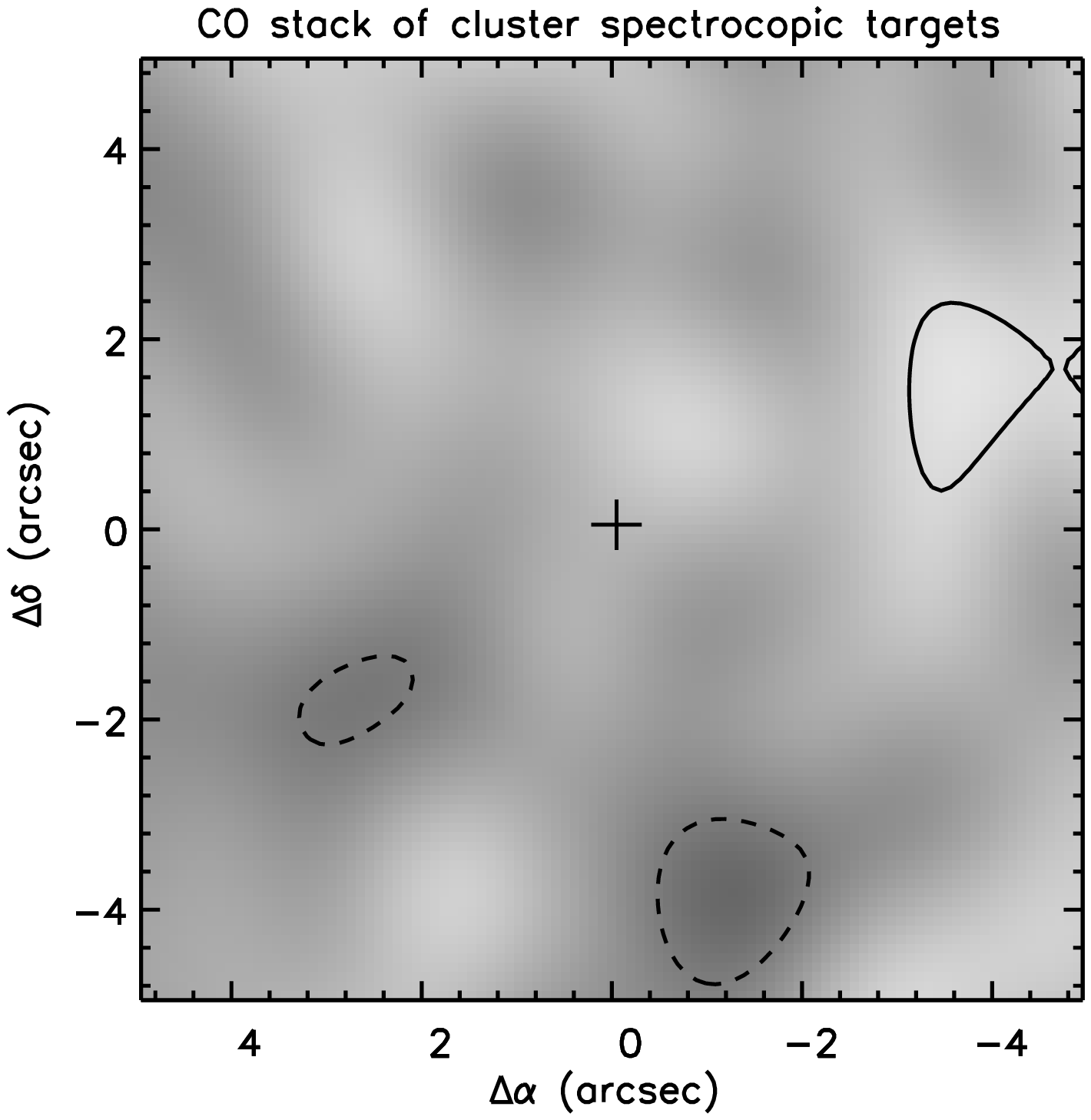}
\includegraphics[scale=0.45]{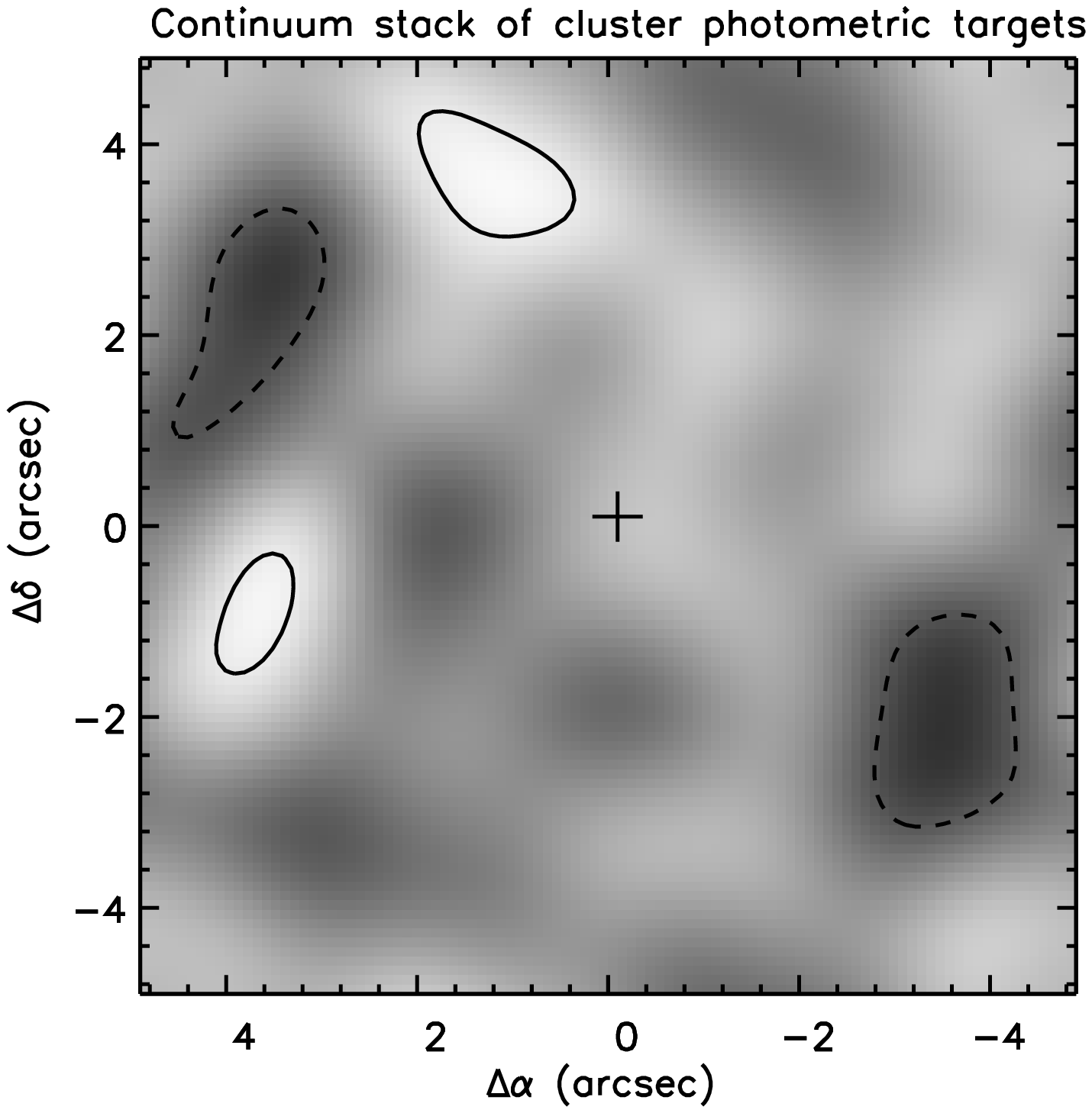}
\caption{{\it Left:} Combined CO emission integrated over 400 km s$^{-1}$ from the six galaxies that have a spectroscopic redshift measurement. Contours represent the flux density in steps of $1\sigma=0.27$ mJy beam$^{-1}$ starting at $\pm2\sigma$. {\it Right:} Combined 3.3 mm continuum emission over 23 galaxies with photometric redshifts in the range $z=0.25-0.35$ within 30\arcsec from the center of the field. Contours represent the flux density in steps of $1\sigma=58\ \mu$Jy beam$^{-1}$ starting at $\pm2\sigma$.}
\end{figure*}

In order to test how the gas is cooling onto the massive galaxies in this galaxy cluster, we mapped the distribution of galaxies, which should roughly trace the filamentary structure and the path of intergalactic gas into the cluster core. We selected galaxies from the SDSS with photometric redshift redshifts between 0.25 and 0.35 and brighter than $i'=23.0$ in an area of $\sim5.2'\times5.2'$ centered in H1821+643. A density map from the selected galaxies was constructed by creating a grid of cells and computing the number density as the 10th nearest neighbor galaxy to each cell position. Figure \ref{fig:density} shows the density map highlighting the position of the CO detected objects. This map is given in terms of the average density, $\delta_0=0.81$ arcmin$^{-2}$, found in a control field to the south of the H1821+643 field, and thus represents the degree of overdensity ($\delta/\delta_0$) with respect to this average. 

H1821+643 is well within the densest part of the cluster. Figure \ref{fig:density} suggests that there is density sub-structure associated with H1821+643, with the stronger density peaks following across the south-west, south-east and north-east of this source. This can also be seen in the distribution of galaxies shown in Fig. \ref{fig:mosaic}. 

To investigate whether substantial amounts of gas have been deposited and condensed in molecular form onto other galaxies in the cluster, we performed a stacking analysis of the CO emission using the galaxies that have accurate spectroscopic redshift measurements. For this, we selected galaxies that lie within a distance of $40''$ from of our target fields central pointing and in redshift range covered by the observing frequency of our observations, $z=0.2925-0.3015$, from the spectroscopic samples of \citet{Tripp1998} and references therein. With this criteria only six galaxies were selected, most of them lying in the outskirts of the image where the sensitivity decreases, as shown in Fig. \ref{fig:mosaic}. We extracted 400 km s$^{-1}$ single-channel maps centered at the position of each galaxy and with a central frequency corresponding to the galaxies redshift (e.g. $\nu_\mathrm{CO}/(1+z)$). We deconvolved the visibilities and the resulting dirty maps were thus stacked to compute a noise-weighted average map. We did not detect individual sources or their combined emission (Fig. 4). We put a $3\sigma$ limit for their average emission of 0.81 mJy per 400 km s$^{-1}$, which constrains the gas mass contained on average in this galaxies to $<1.1\times10^9$ M$_\odot$. This is consistent with the molecular gas masses found in elliptical and lenticular galaxies in the local volume, which range between $2\times10^7-5\times10^9$ M$_\odot$ \citep{Crocker2011}, and excludes the possibility that gas-rich spirals, which typically have molecular gas masses of $>1\times10^9$ M$_\odot$, are in the inner regions of the cluster.

To check whether any possible bright continuum source could be contributing to the bright emission seen in FIR images, we take advantage of the comparatively high-resolution of the 3.3 mm images to search for continuum emission from the galaxies in this cluster field by using stacking. To obtain better statistics, we increased our sample of galaxies in the cluster by selecting sources from the SDSS photometric catalog with redshifts in the range $z=0.25-0.35$, optical emission $i'_\mathrm{AB}<23$ and located within $30''$ from the central position of each target field. This resulted in a sample of 23 galaxies. We combined all line free bands making a total bandwidth of 4.125 GHz at $\sim91$ GHz. Similar to the CO maps, we extracted continuum maps  and deconvolved the visibilities and combined the dirty maps, which were thus stacked in a noise weighted fashion. Again no significant detection was found in the average image down to a 3$\sigma$ limit of 175 $\mu$Jy. This implies that the only strong continuum emitter in the field is indeed H1821+643.

\section{Conclusions}

We have presented the detection of $^{12}$CO $J=1-0$ line emission towards the template ULIRG-to-QSO transition object [H89]1821+643. The extended CO emission is not associated with the QSO or host galaxies but is offset by about 8.8 kpc to the south-east. The molecular gas reservoir is estimated to be $\sim8\times10^9$ M$_\odot$. The CO emission coincides with the position of an optical feature suggested by HST optical imaging. We argue that H1821+643 could either represent a merger between a gas-poor, giant elliptical and a gas-rich companion galaxy or that the CO emission could arise from a tail-like gas-rich structure within the giant elliptical galaxy. 

Based on the optical emission, H1821+643 was found to have a total mass of $\sim2.1\times10^{12}$ M$_\odot$. However, we find that the dynamical mass found for the CO source (offset from the elliptical host) based on its line profile can amount up to $80\%$ of the mass of the elliptical host. On one hand, this suggest that the CO source should be a different galaxy that the elliptical one because otherwise its large mass would imply an highly asymmetric stellar distribution for the elliptical host. On the other hand, this results suggest that if we had measured the elliptical dynamical mass based on the CO profile, assuming it is well centered in the elliptical, we would have underestimated its value.

We revised the dual nature of the SED of H1821+643  (starburst and QSO), and thus associate a cold dust ($\sim50$ K), massive starburst component with the CO emitting source, and the warm dust ($\sim$130 K), AGN component with the nuclei and host galaxy. We find that the inclusion of a flat radio spectrum component, important even up to $\sim350$ GHz, implies a higher temperature for the cold dust component respect to previous estimates.

Based on the projected density map of galaxies in the surrounding cluster field, we find that the field shows an overdensity peak to the south-east of the H1821+643 system, arguing in favor of cold flows of gas toward the inner regions of the cluster along this path. This hypothesis needs to be confirmed with sensitive CO and HI observations, in order to reveal the nature of the CO emitting source and to directly measure the total amount of neutral gas deposited in the central regions of the cluster and QSO. 

The stacking technique was used to estimate the combined CO emission from galaxies in the field with spectroscopic redshifts. This allowed us to put an upper limit to the molecular gas mass of $1.1\times10^9$ M$_\odot$, comparable to that of typical elliptical galaxies. Additionally, the stacking analysis was used to search for 93 GHz continuum emission from a sample of photometrically-selected sources with redshifts in the range between $0.25-0.35$. We found no strong continuum emitters down to a $3\sigma$ limit of 175 $\mu$Jy. The lack of significant amounts of gas in cluster field galaxies hints at the importance of interactions in the central object that can trigger the gas condensation and star formation.

The results exposed above, in which the bulk of the molecular gas is well off the central QSO exemplify the importance of performing a revision of the local ULIRG-to-QSO transition objects with high spatial resolution observations. Many of these objects may have been mis-classified as post merger systems; and perhaps more importantly they are ideal targets to investigate unrelaxed gas-star configurations in which dynamical masses may have been underestimated.

\acknowledgments
M.A. thanks Dominik Riechers and Nikolas Volgenau for their help on setting up the CARMA observations. Support for CARMA construction was derived from the Gordon and Betty Moore Foundation, the Kenneth T. and Eileen L. Norris Foundation, the James S. McDonnell Foundation, the Associates of the California Institute of Technology, the University of Chicago, the states of California, Illinois, and Maryland, and the National Science Foundation. Ongoing CARMA development and operations are supported by the National Science Foundation under a cooperative agreement, and by the CARMA partner universities.

{\it Facilities:} \facility{CARMA}, \facility{HST}.



\clearpage


\clearpage

\end{document}